# Bridge the Gap between Past and Future: Siamese Model Optimization for Context-Aware Document Ranking

Songhao Wu*
Gaoling School of Artificial
Intelligence, Renmin University of
China
Beijing, China
songhaowu@ruc.edu.cn

Quan Tu*
Gaoling School of Artificial
Intelligence, Renmin University of
China
Beijing, China
quantu@ruc.edu.cn

Mingjie Zhong*
Hong Liu
Ant Group
Hangzhou, China
mingjie.zmj@antgroup.com
yizhou.lh@antgroup.com

Jia Xu
Ant Group
Hangzhou, China
steve.xuj@antgroup.com

Jinjie Gu
Ant Group
Hangzhou, China
jinjie.gujj@antgroup.com

Rui Yan[†]
Gaoling School of Artificial
Intelligence, Renmin University of
China
Beijing, China
ruiyan@ruc.edu.cn

## ABSTRACT

In the realm of information retrieval, users often engage in multi-turn interactions with search engines to acquire information, leading to the formation of sequences of user feedback behaviors. Leveraging the session context has proven to be beneficial for inferring user search intent and improving document ranking. A multitude of approaches have been proposed to exploit in-session context for improved document ranking. Despite these advances, the limitation of historical session data for capturing evolving user intent remains a challenge. In this work, we explore the integration of future contextual information into the session context to enhance document ranking. We present the siamese model optimization framework, comprising a history-conditioned model and a future-aware model. The former processes only the historical behavior sequence, while the latter integrates both historical and anticipated future behaviors. Both models are trained collaboratively using gold labels and pseudo labels predicted by each other. The history-conditioned model, referred to as **ForeRanker**, progressively learns future-relevant information to enhance ranking during inference using solely historical session. To mitigate inconsistencies during training, we introduce a peer knowledge distillation method employing a dynamic gating mechanism, allowing models to selectively incorporate information. Experimental results on benchmark datasets demonstrate the effectiveness of our ForeRanker, showcasing its superior performance compared to existing methods.

*Equal contribution.
[†]Corresponding author: Rui Yan (ruiyan@ruc.edu.cn).

## 1 INTRODUCTION

To fulfill a complicated information-seeking task with search engine, users tend to submit queries for multiple turns and thus obtain a satisfying result ultimately. Such a sequence of user's feedback behaviors (e.g. issued queries and clicked documents) within the brief interval of time is often denoted as a *search session*. Many studies [1, 2, 32, 36, 42, 43] have shown that the session context can be leveraged to model user's implicit search intent and better rank the candidate documents as a consequence. An increasing number of methods have been proposed to exploit the in-session contextual information in different manners. Earlier studies employed neural architectures to encode the historical behavior into latent representation for better search intent understanding [1–3]. More recently, many works [4, 42, 43] viewed context-aware document ranking as sequential modeling and concatenated all historical activities into the pre-trained transformer encoder, considering the success of the pre-trained language model (e.g. BERT [7]).

Previous studies primarily rely on the user's historical search log to rank documents. These works propose to capture the semantic interactions between the historical session and candidate documents at fine-grained level to facilitate context-aware document ranking. However, the historical session alone may not contain enough contextual information to obtain desired document ranking performance. The history-conditioned techniques excel at capturing the semantic in-session, while they overlook the overall search intent throughout the entire session due to the limited session context. Future activities together with the on-going session co-constructs an enriched search scenario, which exists cues for current intent modeling. Figure 1 illustrates such a scenario within one session. After putting forward a series of product-related queries and clicks, the user turns to an e-commerce website for purchasing MacBook Pro 13" in the end. Faced with the current query (*Macbook Pro 13"*), if it weren't foretold the user's purchasing intent, the search engine possibly assigns a higher ranking score to the product details (*MacBook Pro 13" - Apple Official Site*), and take merely the semantic relevance into consideration. A search session is a real-world interactive process with user engagement. Beyond the semantic flows throughout the session sequence, more informative relevance judgements can be obtained if we could perceive the underlying search intent of the entire session. Recent research [32] finds that sessions with similar intent can be exploited to improve the current document ranking. However, we luckily find that the search intent contained in similar sessions is easily accessed in the user's future behaviors. This finding inspires us to bring the future information into the ongoing session and thus break the bottleneck of the history-conditioned models.

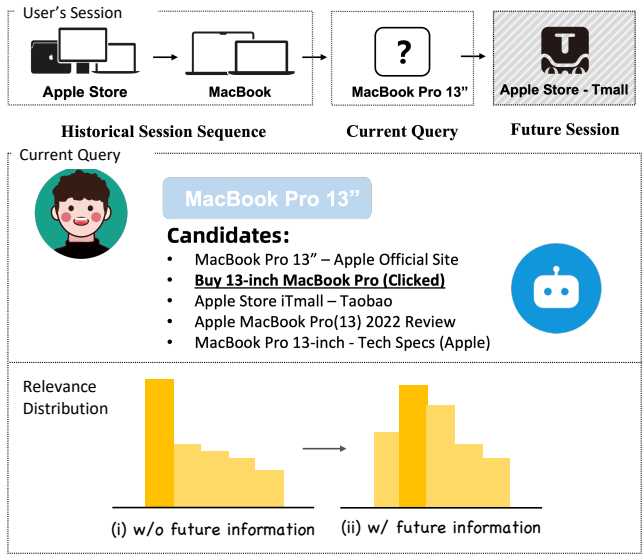

**Figure 1: An example of the session search scenario. Given the prospective user behaviours, the model make relevance judgments that are better aligned with user's current search intent. That demonstrates the potential of modeling the future context in search session.**

Although the future interactions may carry with abundant hints for current search intent understanding, it is rebellious to straightly introduce the prospective contextual information for two reasons: (1) the inaccessibility of future information at inference time. (2) the latent noise contained in future behaviors. Real-world search logs have unavoidable intent shift and it is intractable to filter out these noisy components in future behaviors. To address the issued problems, we propose **siamese model optimization** framework, which consists of a history-conditioned and a future-aware model. The future-aware model is fed with the historical and future session sequence, while the history-conditioned model singly takes the historical as input. Both of the twin models are optimized with the supervised labels and pseudo labels predicted by the other in a peer knowledge distillation paradigm. The history-conditioned model gradually learns the future-heuristic information and ultimately boost better ranking performance at test time, where only the historical behaviors is accessible. To prevent the siamese models from being confused by the inconsistent contextual information during the training process, we devise a peer knowledge distillation method to adaptively filter out the disturbances from each other. It involves an dynamic gating mechanism to selectively block the soft pseudo labels and the gold label by comparision. Further, we incorporate an annealing schedule into our dynamic gating mechanism to gradually transfer our training manner from the supervised learning mode to a mutual learning mode.

Our experimental results demonstrate that the siamese model optimization does relieve the effect of mutual noise during training and effectively improve the resultant history-conditioned model, **ForeRanker**, through future knowledge injection. Comparing all baselines, ForeRanker achieves state-of-the-art performance.

In summary, our contributions in this paper are three folds:

- We propose the siamese model optimization framework, which takes both user's historical and future behaviors into context-aware document ranking. It provides additional contextual constraints to derive a relevance distribution, making it easier to capture the implicit search intent.
- Peer knowledge distillation is introduced to save the siamese models from inconsistent contextual information by adaptively filtering out the disturbances from each other based on the dynamic gating mechanism.
- We conduct comprehensive experiments on two public benchmarks and our ForeRanker shows better results than state-of-the-art baselines on both. Extensive analysis further demonstrates the effectiveness of our proposed siamese model optimization.

## 2 RELATED WORK

### 2.1 Context-aware Document Ranking

It has been demonstrated that the contextual information in sessions makes for user intent modeling. Early studies use statistical language models and heuristic algorithms to extract contextual features from user's search activities [30, 34, 37]. Later, the development of neural networks boosts the emergence of deep learning approaches in context-aware document ranking. For instance, RNN-based architectures are proposed to encode the session sequence. A hidden representation is obtained for document ranking [1–3]. As large-scale pre-trained language model, e.g. BERT [7], has achieved great success in wide range of NLP and IR tasks, it has become the common practice that employing PLM as backbone to model user session [4, 25, 32, 42, 43]. More recently, studies focus more on the proposal of specific training strategy. For example, contrastive learning [42] and curriculum learning [43] have been introduced to improve ranking model optimization. Wang et al. [32] propose a heterogeneous graph-based framework to capture valuable signals in the graph structure. However, the issued approaches above only encode the historical behavior sequence as user modeling. Different from that, our method combines user's future behavior together with the current session sequence, which derive a more informative relevance distribution to guide the history-only document ranking.

### 2.2 Future Modeling

Future information has been verified conducive to some scenarios in NLP tasks [8, 9, 14, 17, 18, 33]. For instance, in dialogue generation [9, 14, 18], researchers exploit the future conversations together with historical utterances to generate more diverse and relevant responses. As to the text generation [28, 33, 41], methods which consider the right side tokens as the "future" for the current token, are proposed to fulfill a target-side augmentation. More recently, future modeling is taken into account in general recommendation [16, 38, 40]. Xu et al. [40] take discriminative but not available features when training to improve online performance. Xie et al. [38] adopt a GAN framework to take advantage of useful future information in recommendation. Through a discriminator that can make use of the future features, the generator get sufficient training and then provide better recommendation as a consequence.

Few studies have considered user's future search activity when it comes to the context document ranking. Some works [1, 2, 4, 6] take the succeeding query and document as generative targets considering that the generative task can generalize the ranking model as reported [13]. In this work, we use the future behavior sequence to construct a search scenario where both historical interactions and future context are taken into account. We demonstrate that the introduction of future-aware setting can regularize model training and help history-conditioned model perform better in document ranking. Moreover, a siamese model optimization is devised to bridge the gap and free our model from the noise in different contexts.

## 3 METHODOLOGY

### 3.1 Task Formulation

Before introducing our method, we first state some notations and briefly formulate the task. We denote the historical queries of user's search session as $Q = \{q_1, q_2, \ldots, q_M\}$, where $M$ is the history length. Each query is represented as the original text that the user submitted to the search engine and has been ordered by their issued timestamp. Given the query $q_i$, its candidate document list is denoted as $D_i = \{d_{i,1}, d_{i,2}, \ldots, d_{i,N}\}$, where $d_{i,j}$ is represented by the text content and its binary click label $y_{i,j}$ ($y_{i,j} = 1$ if the user has clicked this document). Following [4, 32, 42, 43], the issued query together with its first clicked document is used to compose the user's behavior sequence. When user is submitting the $i$-th query, the historical behaviors and future behaviors are referred as $H_i = \{q_1, d_1^+, q_2, d_2^+, \ldots, q_{i-1}, d_{i-1}^+\}$ and $F_i = \{q_{i+1}, d_{i+1}^+, q_{i+2}, d_{i+2}^+, \ldots, q_M, d_M^+\}$ respectively. With the notations above, context-aware document ranking task aims at re-ranking the candidate document set $D_i$ considering the historical behaviors $H_i$ and the issued query $q_i$. In this paper, we further incorporate user's future search activities $F_i$ into the session context to learn better ranking when training.

### 3.2 Overview

The overall framework of our proposed siamese model optimization is illustrated in Figure 2. The siamese model consist a history-conditioned model and a future-aware model. The former models the historical behaviors while the latter models both historical and future behaviors. They are jointly trained to rank the candidate documents at training time. In order to transfer the contextual information hidden in the future behaviors to the history-conditioned model, we devise a novel siamese model optimization approach, which saves the twin models from the noise of the inconsistent context. Finally, the history-conditioned model (referred as **ForeRanker**), improved by future-heuristic information, is used to rank the candidates at inference time. In order to distinguish, we denote the future-aware model as **ForeRanker w. future** respectively.

### 3.3 Session Sequence Modeling

In this section, we introduce the modeling of the session sequences. As shown in the middle part of Figure 2, we use the pre-trained language model, e.g BERT [7], as backbone. Following previous works [4, 25, 42, 43], we concatenate the user behaviors into a sequence, and make the output representation of [CLS] token go through a MLP to calculate the ranking score. Given the query $q_i$ and a candidate document $d_{i,j}$, the ranking score $s(q_i, d_{i,j})$ is calculated by:

$$s(q_i, d_{i,j}) = \text{MLP}(\text{BERT}(X)_{[\text{CLS}]}),\qquad(1)$$

where $X$ is the input to BERT. In case of ForeRanker, the input is:

$$X_h = [[\text{CLS}], H_i, q_i, [\text{SEP}], d_{i,j}, [\text{SEP}]],\qquad(2)$$

which only considers the historical behaviors as the context. In case of the future-aware model, the input is:

$$X_f = [[\text{CLS}], H_i, q_i, [\text{SEP}], d_{i,j}, [\text{SEP}], F_i, [\text{SEP}]],\qquad(3)$$

which considers both the historical behaviors and future behaviors as the context.

Generally, we minimize the negative log likelihood loss to optimize model and learn to rank as follows:

$$\mathcal{L}_{NLL}(\Theta) = -\log \frac{e^{s(q_i, d_i^+)}}{\sum_{d_{i,j} \in D_i} e^{s(q_i, d_{i,j})}},\qquad(4)$$

where $\Theta$ is the model parameters, $d_i^+$ is the first clicked document of issued query $q_i$, $d_{i,j}$ is the document in the candidates $D_i$ and the score function $s(\cdot)$ is defined in Equation 1.

### 3.4 Future Knowledge Injection

Future behaviors have additional session context. The model could better align with the user's underlying search intent, taking both the historical and future behaviors into consideration. However, the future behaviors are invisible at inference time and hard to estimate. To address, we attempt to distill the future knowledge from the future-aware model to the history-conditioned ForeRanker when training, and we hope that the ForeRanker could perform better after the injection of future features.

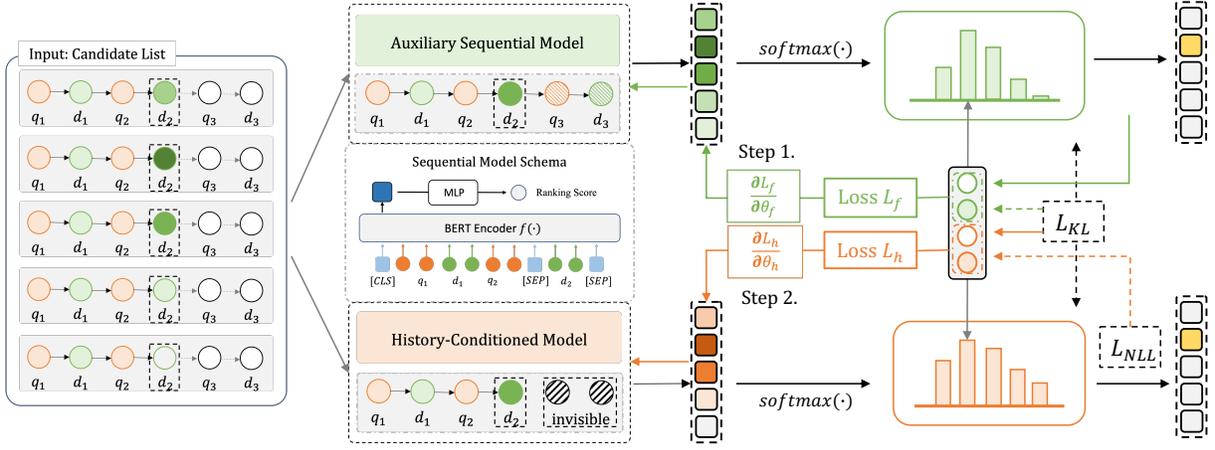

**Figure 2: The overview of the proposed siamese model optimization framework.**

To be specific, we minimize the Kullback-Leibler (KL) divergence [12] between the relevance distributions predicted by the siamese models. Assuming that the parameter of ForeRanker is $\Theta_h$ and the parameter of future-aware model is $\Theta_f$. Given the query $q_i$ and document $d_{i,j}$, the ranking scores of the siamese models are denoted as $s_h(q_i, d_{i,j})$ and $s_f(q_i, d_{i,j})$ respectively.[1] We yield the predicted distributions by apply the softmax function on the ranking scores of candidate documents $D_i$ as follows:

$$\tilde{s}_h(q_i, d_{i,j}) = \frac{e^{s_h(q_i, d_{i,j})}}{\sum_{d_{i,k} \in D_i} e^{s_h(q_i, d_{i,k})}}, \qquad (5)$$

$$\tilde{s}_f(q_i, d_{i,j}) = \frac{e^{s_f(q_i, d_{i,j})}}{\sum_{d_{i,k} \in D_i} e^{s_f(q_i, d_{i,k})}}, \qquad (6)$$

where both $s_h(q_i, d_{i,j})$ and $s_f(q_i, d_{i,j})$ can be calculated by Equation 1. The difference is that the input for $s_h(q_i, d_{i,j})$ is from Equation 2 while the the input for $s_f(q_i, d_{i,j})$ is from Equation 3.

At each training step, the future-aware model estimates a pseudo relevance distribution, considering both the historical and future behaviors. Then we make the predictions of the history-conditioned ForeRanker to approximate the pseudo label by minimizing the KL-divergence between the two distributions to inject the future knowledge:

$$\mathcal{L}_{KL}(\Theta_h) = \sum_{d_{i,j} \in D_i} \tilde{s}_f(q_i, d_{i,j}) \cdot \log \frac{\tilde{s}_f(q_i, d_{i,j})}{\tilde{s}_h(q_i, d_{i,j})}. \qquad (7)$$

### 3.5 Siamese Model Optimization

It is quite conceivable to relate the *future knowledge injection* to the existing knowledge distillation methods, which take the well-trained future-aware model as the *teacher model* and the history-conditioned model as the *student model*. However, due to the unavoidable shift of the user intent in real-world, the noisy supervised signal of the teacher model may mislead the student model and make it confused at inference time. Considering that, it is unwise to simply inject the future knowledge based on the traditional knowledge distillation methods. Experimental analysis in Section 5.2 further demonstrates this viewpoint.

In view of the *noise* in users' future behaviors, we propose a novel **siamese model optimization** framework to facilitate more effective knowledge transfer for the future. Taken the two models (future-aware model and history-conditioned model) as the siamese models, a peer knowledge distillation method is introduced to bridge the gap between past and future and denoise the two distinct search context respectively.

#### 3.5.1 Peer Knowledge Distillation.
Even though the future-aware model introduces extra behaviors as reference, not all queries in-session benefit from the future information. Thus, employing the well-trained future-aware model to perform knowledge distillation may not be the optimal choice. Unlike the conventional knowledge distillation paradigm, where knowledge is transferred from *teacher model* to *student model*, we consider the siamese models, the future-aware model and history-conditioned ForeRanker as equals. They share the same model architecture, initialized parameters and are jointly optimized during training. The distinction between them simply lies in the input: the former receives only the users' historical behaviors, while the latter receives both the user's historical and future behaviors.

Peer knowledge distillation no longer fixes one of the models as the role of *teacher* and the other as *student*. At each training step, we adaptively change the teacher-student status of the siamese models. As we distill the future knowledge from $\tilde{s}_f$ to $\tilde{s}_h$ (Equation 7), we introduce an inverse distillation (Equation 8) as dual task to optimize the future-aware model:

$$\mathcal{L}_{KL}(\Theta_f) = \sum_{d_{i,j} \in D_i} \tilde{s}_h(q_i, d_{i,j}) \cdot \log \frac{\tilde{s}_h(q_i, d_{i,j})}{\tilde{s}_f(q_i, d_{i,j})}, \qquad (8)$$

thus, both siamese models are jointly optimized with the pseudo label ($\mathcal{L}_{KL}$) and the gold label ($\mathcal{L}_{NLL}$) as supervised signal.

---
[1]Here, $h$ stands for "history", and $f$ represents "future".

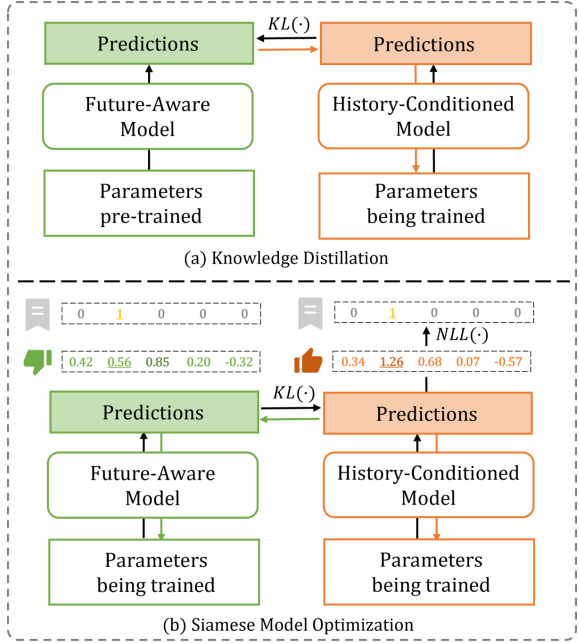

**Figure 3: (a) The conventional knowledge distillation paradigm. The pre-trained future-aware model is considered as the teacher model. Future information is distilled to the history-conditioned model by minimizing a KL term. (b) Siamese model optimization with peer knowledge distillation and dynamic gating mechanism. Given the predictions of the two models, we adaptively adjust the roles of the siamese models and dynamically adopt the training loss.**

Other than blindly following the predictions given by the future-aware model, the peer knowledge distillation allows the history-conditioned model to learn adaptively. Besides, the future-aware model could benefit from the feedback of the history-conditioned ones. The major difference of the peer knowledge distillation and the traditional framework lies in two aspects: 1) There exist no fixed *teacher model* and *student model* in our proposed peer knowledge distillation. 2) The siamese models have the same scale of model parameters and are jointly trained.

*3.5.2 Dynamic Gating Mechanism.* Now with the history-conditioned model $\Theta_h$ and future-aware model $\Theta_f$ at hand, we further propose a dynamic gating mechanism to gradually ensemble the siamese models to achieve a collaborative optimization.

As aforementioned, the siamese models are adaptively considered as teacher model and student model. At each training step, we provide the ground-truth guidance for the temporary teacher model. Specifically we adopt the negative log likelihood loss in Equation 4 as the supervision. Meanwhile we minimize the KL-divergence between the two distributions as in Equation 7 and 8, to distill the peer guidance from the teacher to the student.

To keep the *student model* off from the error accumulation by *teacher model*'s false predictions, we adapatively choose model with better performance as the teacher in training time. As is presented in

Figure 3, if the history-conditioned model (or future-aware model) assigns a higher relevance score $s(q_i, d_{i,j})$ to the positive document than the other, we simply take it as the teacher and the other as student. We optimize *teacher model* with the negative log likelihood loss and the student with the KL-divergence. The ultimate optimization objective for the siamese models are formulated as:

$$\mathcal{L}(\Theta_f) = \alpha \cdot \mathcal{L}_{NLL}(\Theta_f) + (1 - \alpha) \cdot \mathcal{L}_{KL}(\Theta_f), \quad (9)$$

$$\mathcal{L}(\Theta_h) = (1 - \alpha) \cdot \mathcal{L}_{NLL}(\Theta_h) + \alpha \cdot \mathcal{L}_{KL}(\Theta_h), \quad (10)$$

where the signal $\alpha$ is a 0-1 indicator: $\alpha = 1$ if the future-aware model assigns a higher relevance score to the positive candidate, otherwise $\alpha = 0$. Formally, $\alpha$ is defined as:

$$\alpha = \mathcal{I}\left(s_f(q_i, d_i^+) > s_h(q_i, d_i^+)\right). \quad (11)$$

*3.5.3 Training.* For the overall training, we set up a warm-up schedule to control the siamese model optimization. As is previously noted, neither of the two models is well-trained and starts training jointly from scratch. Considering that, the original pseudo labels given by the peer model may be suboptimal. As both models are still underfitting, their predictions may affect the training of another. Hence, an auxiliary ground-truth guidance is provided with the student model during the warm-up schedule. Let $\omega$ denotes the balanced weight of the losses $\mathcal{L}_{NLL}$ and $\mathcal{L}_{KL}$. We gradually decrease $\omega$ from 1 to 0 with a polynomial decay schedule to accelerate training. Once the siamese models have been warmed up, the student model turns to train with the $\mathcal{L}_{KL}$ only. At inference time, only the well-trained history-conditioned is used.

## 4 EXPERIMENTAL SETUP

### 4.1 Datasets and Evaluation Metrics

*4.1.1 Datasets.* We conduct our experiments on two public benchmark datasets, AOL Search log [23] and Tiangong-ST search log [5], which are widely used in earlier context-aware document ranking works [3, 4, 42, 44].

**AOL** search log is a large-scale search log, which contains sessions made up of user's consecutive queries in real-world search scenarios. For fair comparison, we use the one provided by Ahmad et al. [2], where candidates are retrieved by the BM25 algorithm [27]. Each query of the training and validation sets has 5 candidate documents and 50 candidates are provided as to the test set. More details are referred to [1].

**Tiangong-ST** is collected from a Chinese commercial search engine. It has a search log for 18 days and each query is provided with 10 candidate documents. In training and validation setting, the click-through labels are given as implicit relevance feedback. For test set, the last query of each session has a manually annotated relevance score, while the others within the session have only click labels. To keep the consistency, we only use the queries with click signals as test data.

The statistics of the two datasets are presented in Table 1.

*4.1.2 Evaluation Metrics.* Similar to the earlier studies [3, 4, 32, 42, 44], we adopt the Mean Average Precision (MAP), Mean Reciprocal Rank (MRR), and Normalized Discounted Cumulative Gain (NDCG) at position $k$ (NDCG@$k$, $k = 1, 3, 5, 10$) as metrics. Evaluation results are computed by the TREC's official evaluation tool (trec_eval) [10].

**Table 1: The statistics of the two datasets used in our paper. The number in parentheses is the average number of relevant documents.**

| AOL | Training | Validation | Test |
|---|---|---|---|
| # Sessions | 219,748 | 34,090 | 29,369 |
| # Queries | 566,967 | 88,021 | 76,159 |
| Avg.# Query per Session | 2.58 | 2.58 | 2.59 |
| # Candidate per Query | 5 | 5 | 50 |
| Avg. Query Len | 2.86 | 2.85 | 2.9 |
| Avg. Document Len | 7.27 | 7.29 | 7.08 |
| Avg. # Clicks per Query | 1.08 | 1.08 | 1.11 |
| **Tiangong-ST** | **Training** | **Validation** | **Test** |
| # Sessions | 143,155 | 2,000 | 2,000 |
| # Queries | 344,806 | 5,026 | 6,420 |
| Avg.# Query per Session | 2.41 | 2.51 | 3.21 |
| # Candidate per Query | 10 | 10 | 10 |
| Avg. Query Len | 2.89 | 1.83 | 3.46 |
| Avg. Document Len | 8.25 | 6.99 | 9.18 |
| Avg. # Clicks per Query | 0.94 | 0.53 | (3.65) |

## 4.2 Baseline

We compare our methods with two kinds of baselines in our experiment, including ad-hoc ranking methods and context-aware ranking methods.

*4.2.1 Ad-hoc ranking methods.* Ad-hoc ranking methods take no accounts of the session context and focus simply on the semantic matching between the current query and the candidate. All baseline methods follow the same setup as well as our ForeRanker and have been fine-tuned on the evaluating dataset if needed.

• **BM25** [27] is a sparse retrieval algorithm to estimate the relevance of documents to the query. It calculates the relevance scores by considering term frequency and inverse document frequency respectively.

• **ARC-I** [11] is a representation-based method, where convolutional neural networks (CNNs) are used to represent the query and document respectively. The ranking score is given by a multi-layer perception (MLP) based on the representations .

• **ARC-II** [11] is an interaction-based method. Matching map is built considering the word-level interaction. 2-D CNNs are employed to capture semantic relevance features.

• **KRNM** [39] is another interaction method that performs fine-grained interactions between query and document. The ranking score is obtained by the kernel pooling on the matching matrix.

• **Duet** [19] extracts interaction-based and representation-based ranking features by layers of CNNs and MLPs, the ranking score is given taking both these two features in consideration.

• **monoBERT** [21] is a cross-encoder re-ranker based on BERT. It is one of the early attempts to utilize a pre-trained language model for document ranking.

• **monoT5** [22] is a sequence-to-sequence re-ranker that leverages T5 [26] to calculate the relevance score. The T5 model is fine-tuned with a text-generation loss for relevance ranking.

*4.2.2 Context-aware ranking methods.* Context-aware ranking methods leverage both user's historical interactions and the current query to rank the candidate documents.

• **M-NSRF** [1] is a multi-task model that jointly train the document ranking with the query suggestion task. It encodes the historical behavior sequence into latent representation and calculate the ranking scores based on the query, the history and the document.

• **CARS** [2] shares the same tasks with M-NSRF. Different from M-NSRF, CARS additionally models the historical clicked documents through RNN. Attention mechanism is applied for better representation of the session sequence.

• **HBA** [25] concatenates all previous query, clicked documents and the skipped documents as a long sequence into the Transformer architecture with specifically-designed behavior embedding and relative position embedding. The ranking score is computed based on the representation of the "[CLS]" token similar to our work.

• **COCA** [42] pre-train a BERT encoder with data augmentation and contrastive learning to acquire better session representation.

• **ASE** [4] designs three generative tasks to help the encoding of the current session sequence. In contrast to other multi-task approaches that consider only the subsequent query generation(query suggestion) as the auxiliary task, it further take the succeeding clicked document and a similar query as generation targets.

• **HEXA** [32] constructs two heterogeneous graphs, a session graph and a query graph, to capture user intent from global and local aspects respectively. It is the first attempt to go beyond the ongoing search session with a graph structure in recent context-aware document ranking research.

## 4.3 Implementation Details

We implement our model with PyTorch [24], and the BERT model checkpoint is provided by HuggingFace [35]. Specifically, for the capacity of the future information, we consider the subsequent 2 turns of interactions as our future information. Considering that there exist some queries without clicked document in the subsequent interactions in Tiangong-ST, we neglect these query-document pairs and replace them with the similar query and a corrupted clicked document, inspired by [4, 42] as the pseudo-future. The BERT, used for sequential modeling, is initialized with the original parameters from HuggingFace. We adopt AdamW optimizer [15] to train our model. The learning rate is set as 2e-5 with a linear decay. We train our method by five epochs and we set different batch sizes on two datasets as they have different numbers of candidate documents. More details can be found in our code[2].

## 5 RESULTS AND ANALYSIS

### 5.1 Overall Results

The overall results are shown in Table 2. Our proposed model ForeRanker performs better than all of the existing baselines, which further verifes the superiority of our approach. On the basis of that, we make the following observations in addition:

(1) **ForeRanker outperforms all baselines in terms of all metrics on both datasets, which demonstrates the effectiveness of introducing future features into context-aware document ranking.** The proposed training schema bridges the gap between the past and future, allowing the history-conditioned model to peep into the potential future behaviors and thereby enhancing better understanding of current user intent. Our method significantly outperforms a series of methods that simply take user's historical behaviours as input to improve ranking performance. The improvement of our ForeRanker suggests that the practice to utilize the future features is promising, which is more conducive to user intent modeling and helps with better document ranking.

---

[2]https://github.com/wsh-rucgl/ForeRanker

**Table 2: The overall results of our model and compared baselines on two datasets. The best performance and the second best performance are in bold and underlined respectively. The line depicting the performance of ForeRanker w. future is in grey, as the future behavior is inaccessible during inference. "†" and "‡" indicate our model achieves significant improvements over all existing methods in paired t-test with p-value < 0.01 and p-value < 0.05 respectively (with Bonferroni correction).**

| Model | AOL | | | | | | Tiangong-ST Click | | | | | |
|---|---|---|---|---|---|---|---|---|---|---|---|---|
| | MAP | MRR | NDCG@1 | NDCG@3 | NDCG@5 | NDCG@10 | MAP | MRR | NDCG@1 | NDCG@3 | NDCG@5 | NDCG@10 |
| BM25 | 0.2200 | 0.2271 | 0.1195 | 0.1862 | 0.2136 | 0.2481 | 0.2845 | 0.2997 | 0.1475 | 0.1983 | 0.2447 | 0.4527 |
| ARC-I | 0.3361 | 0.3475 | 0.1988 | 0.3108 | 0.3489 | 0.3953 | 0.6597 | 0.6826 | 0.5315 | 0.6383 | 0.6946 | 0.7509 |
| ARC-II | 0.3834 | 0.3951 | 0.2428 | 0.3564 | 0.4026 | 0.4486 | 0.6729 | 0.6954 | 0.5458 | 0.6553 | 0.7086 | 0.7608 |
| KRNM | 0.4038 | 0.4133 | 0.2397 | 0.3868 | 0.4322 | 0.4761 | 0.6551 | 0.6748 | 0.5104 | 0.6415 | 0.6949 | 0.7469 |
| Duet | 0.4008 | 0.4111 | 0.2492 | 0.3822 | 0.4246 | 0.4675 | 0.6745 | 0.7026 | 0.5738 | 0.6511 | 0.6955 | 0.7621 |
| monoBERT | 0.5499 | 0.5607 | 0.3978 | 0.5518 | 0.5850 | 0.6151 | 0.7284 | 0.7566 | 0.6352 | 0.7129 | 0.7566 | 0.8040 |
| monoT5 | 0.5464 | 0.5571 | 0.3948 | 0.5476 | 0.5810 | 0.6102 | 0.7215 | 0.7505 | 0.6239 | 0.7084 | 0.7513 | 0.7991 |
| M-NSRF | 0.4217 | 0.4326 | 0.2737 | 0.4025 | 0.4458 | 0.4886 | 0.6836 | 0.7065 | 0.5609 | 0.6698 | 0.7188 | 0.7691 |
| CARS | 0.4297 | 0.4408 | 0.2816 | 0.4117 | 0.4542 | 0.4971 | 0.6909 | 0.7134 | 0.5677 | 0.6764 | 0.7271 | 0.7746 |
| HBA | 0.5281 | 0.5384 | 0.3773 | 0.5241 | 0.5624 | 0.5951 | 0.6957 | 0.7171 | 0.5726 | 0.6807 | 0.7292 | 0.7781 |
| COCA | 0.5500 | 0.5601 | 0.4024 | 0.5478 | 0.5849 | 0.6160 | <u>0.7481</u> | <u>0.7696</u> | <u>0.6386</u> | <u>0.7445</u> | <u>0.7858</u> | <u>0.8180</u> |
| ASE | <u>0.5650</u> | <u>0.5752</u> | <u>0.4144</u> | <u>0.5682</u> | <u>0.6007</u> | <u>0.6283</u> | 0.7410 | 0.7637 | 0.6277 | 0.7381 | 0.7790 | 0.8130 |
| HEXA | 0.5625 | 0.5727 | 0.4142 | 0.5631 | 0.5974 | 0.6279 | 0.7427 | 0.7660 | 0.6352 | 0.7378 | 0.7790 | 0.8141 |
| ForeRanker | **0.5737**† | **0.5841**† | **0.4255**† | **0.5772**† | **0.6097**† | **0.6369**† | **0.7537**‡ | **0.7752**‡ | **0.6503** | **0.7487**‡ | **0.7880**‡ | **0.8220**‡ |
| w. future | 0.5798 | 0.5899 | 0.4311 | 0.5845 | 0.6159 | 0.6429 | 0.7781 | 0.8034 | 0.6873 | 0.7779 | 0.8119 | 0.8415 |

**Table 3: The results of the ablation analysis.**

| | w/o future | | w/o peer | | w/o gating | | **ForeRanker** |
|---|---|---|---|---|---|---|---|
| MAP | 0.5574 | -2.84% | 0.5574 | -2.84% | 0.5681 | -0.97% | **0.5737** |
| MRR | 0.5678 | -2.79% | 0.5676 | -2.82% | 0.5782 | -1.01% | **0.5841** |
| NDCG@1 | 0.4078 | -4.16% | 0.4071 | -4.32% | 0.4196 | -1.39% | **0.4255** |
| NDCG@3 | 0.5586 | -3.23% | 0.5591 | -3.14% | 0.5702 | -1.21% | **0.5772** |
| NDCG@5 | 0.5923 | -2.85% | 0.5923 | -2.85% | 0.6036 | -1.00% | **0.6097** |
| NDCG@10 | 0.6218 | -2.37% | 0.6212 | -2.47% | 0.6317 | -0.82% | **0.6369** |

(2) **ForeRanker can effectively utilize the future information with our proposed siamese model optimization.** As the future search activities is inaccessible, our ForeRanker model achieves a worse, yet competitive ranking performance compared with its future-aware siamese model. The relatively low performance of ForeRanker w. future further verified the existence of noise in session future. We further demonstrate the effectiveness of our siamese model optimization in the ablated analysis in Section 5.2.

(3) It is noticed that the improvement of our model on AOL dataset is more significant than that on Tiangong-ST. Similar observation can be found in the earlier studies [32]. As is shown in Table 5.1, existing works fail to perform consistently on the two datasets. One possible explanation is that the candidate documents in Tiangong-ST are collected from the search log of a Chinese commercial search engine. It might be intractable for existing approaches to distinguish the differences between the candidates in real-world search engine and thus limits their application in real-world scenarios. However, our ForeRanker performs best on both, which demonstrates its potential for application in existing industrial search engines.

## 5.2 Ablation Study

We make ablation studies from several aspects and investigate the effects of various designs for the proposed siamese model optimization framework. Specifically, we conduct ablation experiments on AOL dataset as follows.

**ForeRanker w/o future** means we don't take the future features into consideration during model training. Instead we solely optimize the history-conditioned model $\Theta_h$ with Equation 4.

**ForeRanker w/o peer** means we reject the proposed peer knowledge distillation setup and distill a well-trained future-aware model into the history-conditioned model, following the traditional knowledge distillation paradigm.

**ForeRanker w/o gating** means we exclude the proposed dynamic gating mechanism. Alongside the ground-truth guidance, the siamese models are mutually updated based on the KL-divergence, with no consideration of the specific model performance.

Table 3 show the performance of the ablated variants. All the ablated models perform worse, and additionally, we can draw the following conclusions:

(1) **The introduction of future information in the siamese model optimization greatly boosts document ranking performance.** Compared to the conventional history-conditioned model that has no access to the prospective search activities, the significant improvement demonstrates the crucial role that the future-aware model has played in the proposed siamese model optimization. The experimental results provide additional support for our assumption that constructing an enriched search scenario with future features indeed favours the capture of user search intent.

(2) **The proposed peer knowledge distillation scheme can make better use of the relevant features in prospective search behaviors.** Forcing the history-conditioned model to imitate the outputs given by the well-trained future-aware model results in an unexpected performance drop (2.8% decrease in terms of MAP), which even comes close to the performance of **ForeRanker w/o future**. As stated in Section 3.5, there is noise in the evolving search scenario that may confuse models' relevance judgement. Peer knowledge distillation makes it possible for the history-conditioned model to distinguish the noise and be saved from the disturbance, thereby facilitating better future knowledge transfer.

**Table 4: Performance of the history-conditioned model using label smoothing.**

|  | MAP | MRR | NDCG@1 | NDCG@3 | NDCG@5 | NDCG@10 |
|---|---|---|---|---|---|---|
| ForeRanker | **0.5737** | **0.5841** | **0.4255** | **0.5772** | **0.6097** | **0.6369** |
| - w/o future, $\alpha = 0$ | 0.5574 | 0.5678 | 0.4078 | 0.5586 | 0.5923 | 0.6218 |
| - w/o future, $\alpha = 0.1$ | 0.5615 | 0.5720 | 0.4116 | 0.5641 | 0.5970 | 0.6254 |
| - w/o future, $\alpha = 0.2$ | 0.5549 | 0.5653 | 0.4029 | 0.5571 | 0.5908 | 0.6198 |

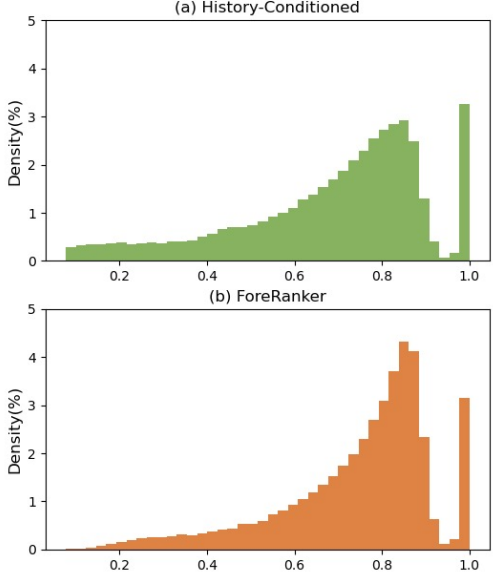

**Figure 4: Distribution of entropy on the test set of AOL dataset. The entropy is calculated with the relevance probability given by the model predictions.**

(3) **The well-designed gating mechanism can bring in additional bonus in model training.** The performance of our ablated model experiences a slight decrease (1.0% decrease in terms of MAP) when we remove the gating mechanism. This aligns with our design intent that, the dynamic gating mechanism can mitigate the negative influence of the peer model's misjudgment and enhance a stable training process as a consequence.

### 5.3 Insight Into Siamese Model Optimization

While the experimental results in Section 5.1 have empirically confirmed the effectiveness of our method, it is important to discuss how our siamese model optimization benefit from the future information to improve ranking performance. To clarify that, we conduct further analysis of our ForeRanker by comparing its predictions with the ablated history-conditioned model that has no access to the future information (ForeRanker w/o future in Section 5.2).

Given the two models' predictions, we calculate the entropy [29] of the predicted relevance probability. The probability is obtained by applying the softmax function on the ranking scores as in Equation 5 and 6:

$$\tilde{s}(q_i, d_{i,j}) = \frac{e^{s(q_i, d_{i,j})}}{\sum_{d_k \in D_i} e^{s(q_i, d_{i,k})}},\tag{12}$$

and the entropy can be formulated as:

$$\mathcal{H}(\tilde{s}(q_i)) = -\sum_j \tilde{s}(q_i, d_{i,j}) \log \tilde{s}(q_i, d_{i,j}) = \mathbb{E}[-\log \tilde{s}(q_i, d_{i,j})].\tag{13}$$

Figure 4 visualised the distribution of the entropy on the test set of AOL. The reported results are normalized with the maximum possible entropy to the range of 0 to 1.

Compared to ForeRanker, the distribution of the history-conditioned model is more concentrated in the low entropy region. A small entropy of the relevance probability indicates that the relevance scores assigned are more concentrated on a few candidate documents. This happens when the model is too-confident about its predictions and can lead to the suboptimal performance.

Previous studies propose some regularization methods [20, 33], e.g label smoothing [31], to address the issue. For model trained with label smoothing of parameter $\gamma$, we minimize the cross-entropy between the modified targets $y^{LS}$ and the model outputs $\tilde{s}$:

$$y_{i,j}^{LS} = y_{i,j} \cdot (1 - \gamma) + \frac{\gamma}{|D_i|}\tag{14}$$

The siamese model optimization solve the dilemma with the help of the future information. Given an enriched search context, the predictions of the future-aware model introduce context-aware smoothing factors to regularize ForeRanker. This may account for the superiority of our proposed siamese model optimization. We compare the performance of ForeRanker with history-conditioned model using label smoothing in Table 4 to further verify that.

### 5.4 Effect of the Capacity of Future Information

Specifically, we consider the subsequent $k$ turns of interactions, along with the historical behaviors, to build the future-aware search context, as stated in Section 3.3. The hyperparameter $k$ is chosen based on the performance on the validation set. To demonstrate the impact of the capacity of future information on our model's performance, we additionally conduct training and testing on AOL dataset choosing different values for $k$ respectively. We vary the choice of $k$ from 1 to 3, as the average session length is relatively small (2.5). Therefore it makes no sense to consider too many turns of search activities. The results are shown in Table 5.

In general, models achieve a consistent improvement over the conventional history-conditioned model, which further indicates the robustness of our proposed siamese model optimization in taking full advantage of the future information. From Table 5, we can find that our model experiences a slight performance drop when provided with the subsequent 3 turns of interaction. No further promotion is found even when considering a more informative search scenario. We believe that the too-distant user behaviors in the future may disturb current intent detection, which aligns with our original intuition to denoise the future behavior sequence.

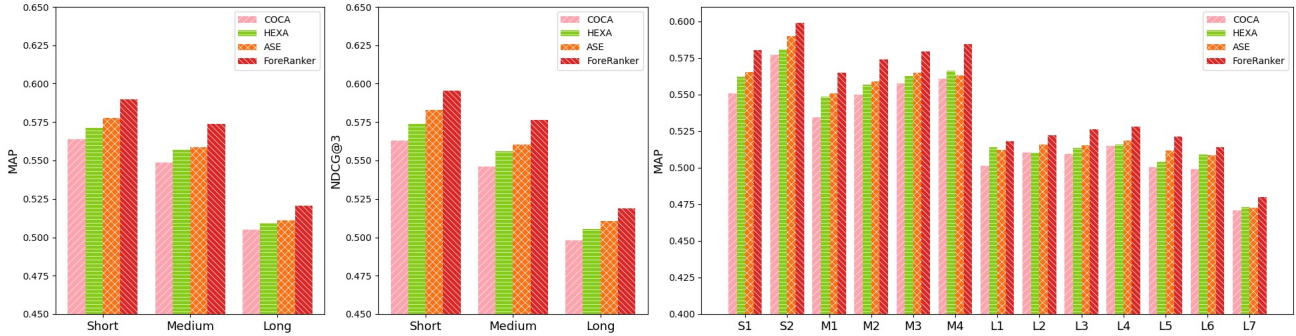

**Figure 5: Performance comparison of COCA, HEXA, ASE and ForeRanker with different session lengths and at different query positions. (Left) Performance with different lengths of sessions. (Right) Performance at different query positions in short (S1-S2), medium (M1-M4), and long sessions (L1-L7). The number after "S", "M", or "L" represents the query index in the session.**

**Table 5: Performance of ForeRanker with different turns of future interactions on AOL dataset.**

|  | MAP | MRR | NDCG@1 | NDCG@3 | NDCG@5 | NDCG@10 |
|---|---|---|---|---|---|---|
| ForeRanker ($k$=1) | 0.5733 | 0.5834 | 0.4250 | 0.5770 | 0.6087 | 0.6359 |
| Improv. | 2.85% | 2.75% | 4.22% | 3.29% | 2.77% | 2.27% |
| ForeRanker ($k$=2) | **0.5737** | **0.5841** | **0.4255** | **0.5772** | **0.6097** | **0.6369** |
| Improv. | **2.92%** | **2.87%** | **4.34%** | **3.33%** | **2.94%** | **2.43%** |
| ForeRanker ($k$=3) | 0.5721 | 0.5822 | 0.4232 | 0.5758 | 0.6079 | 0.6352 |
| Improv. | 2.64% | 2.54% | 3.78% | 3.08% | 2.63% | 2.16% |
| ForeRanker (w/o future) | 0.5574 | 0.5678 | 0.4078 | 0.5586 | 0.5923 | 0.6218 |

## 5.5 Effect of Session Lengths and Query Positions

In this section, we study the performances of our ForeRanker on sessions with different lengths. Following previous works [3, 4, 42–44], we split the test set of AOL dataset into three categories as follows:

(1) Short sessions (with 2 queries) - 66.53% of the test set.
(2) Medium sessions (with 3-4 queries) - 27.24% of the test set.
(3) Long sessions (with 5+ queries) - 6.23% of the test set.

We compare our method with COCA, ASE and HEXA on the AOL dataset. The results regarding MAP and NDCG@3 are presented in the left part of Figure 5. To begin with, our model achieve better results than all of the compared baselines across all three subsets. The robustness of our proposed training scheme boosts consistent performance in the face of search contexts with various session lengths. Secondly, we can observe that our model achieves a distinct improvement over the existing methods in shorter sessions. We attribute it to the utilization of the future information in our model training, which enriches the current session and fosters a better understanding of the user's overall search intent.

To gain insights into our model's performance as the search session progresses, we further compare our model with other approaches at different query positions, as illustrated in the right part of Figure 5. Generally speaking, the context-aware ranking models show gradual performance improvement considering that there is more contextual information to employ as the session progresses. It is noticeable that all models' performances are reported to be worse in long sessions and experience a decrease in the last few queries. This phenomenon can be attributed to the fact that long sessions may involve unexpected shifts in intent or exploratory search tasks. Moreover, obtaining future information about the last queries in long sessions is impossible, which may contribute to the relatively poor performance of our model in such situations.

## 6 CONCLUSION

In this paper, we introduce ForeRanker to addresses the challenges in context-aware document ranking within search sessions. To break the bottleneck of traditional methods, we first bring the future user behaviors into context modeling. A novel siamese model optimization framework is introduced to handle the inconsistencies in contextual information during training, presenting a novel way to incorporate both historical and future user behaviors. Extensive experiments are conducted on two large-scale benchmarks and the results demonstrate the superiority of our ForeRanker. Our work is an early attempt of future modeling in context-aware document ranking and we are moving forward more trials.

## ACKNOWLEDGMENTS

This work was supported by the National Natural Science Foundation of China (NSFC Grant No.62122089), Beijing Outstanding Young Scientist Program NO. BJJWZYJH012019100020098, Intelligent Social Governance Platform, Major Innovation & Planning Interdisciplinary Platform for the "Double-First Class" Initiative, Renmin University of China, the Fundamental Research Funds for the Central Universities, the Research Funds of Renmin University of China, and the Ant Group.